\documentclass[aps,prd,showpacs]{revtex4}

\usepackage{amsmath}
\usepackage{amssymb}
\usepackage{bm}% bold math
\usepackage{graphicx}
\usepackage{multirow}
\usepackage{textcomp}

\allowdisplaybreaks[1]

\begin{document}

\title{\boldmath Deeply inelastic scattering as an inverse problem}
\author{Hsiang-nan Li}
\affiliation{Institute of Physics, Academia Sinica,
Taipei, Taiwan 115, Republic of China}

\date{\today}

\begin{abstract}
A novel formalism for high-energy inclusive QCD processes based on analyticity of a 
scattering amplitude is proposed, which is conceptually distinct from the conventional 
factorization theorem. We illustrate how to analyze proton deeply inelastic scattering  
by means of the dispersion relations for double deeply virtual Compton scattering 
defined on a complex baryon mass plane. Inputting the perturbative results for the 
heavy baryon scattering amplitudes, we solve for the proton structure functions 
$F_2$ and $F_L$ in the nonperturbative low-mass region directly from the dispersion 
relations, i.e., as an inverse problem. The prior information, including the scaling of the 
structure functions in photon virtuality $Q^2$ and the quark counting rules for the threshold 
region with the Bjorken variable $x_B\to 1$, is incorporated. The only nonperturbative 
parameter at leading-power precision in $1/Q$ is the proton mass, with which we demonstrate 
the agreement between our predictions and the data for $F_2$ and $F_L$ at intermediate 
$x_B=0.18$-0.65 in a wide range of $Q^2=10^2$-$10^4$ GeV$^2$. Parton distribution functions 
can be extracted from the obtained structure functions as a byproduct of the analysis by 
treating the associated factorization formulas as another inverse problem. The same
formalism with the pion mass as the nonperturbative parameter explains the data for the
pion structure function $F_2^\pi$.

\end{abstract}

\maketitle

\section{INTRODUCTION}

An essential ingredient for QCD theories addresses the handling of nonperturbative dynamics. 
A broadly adopted nonperturbative approach is based on QCD sum rules \cite{SVZ}, which are 
constructed from analytical correlation functions for physical observables. Recently,
we proposed to manage QCD sum rules as an inverse problem \cite{Li:2020ejs}, which improves
the control of theoretical uncertainties; a spectral density on the hadron side of a 
dispersion relation, including resonance and continuum contributions, is regarded as 
an unknown, and solved with standard operator-product-expansion (OPE) inputs on the quark 
side. The existence of a resonance is not presumed in a spectral density, the quark-hadron 
duality is not implemented for parametrizing excited-state contributions, and the 
discretionary prescription for designating a stability window in a Borel mass 
\cite{Coriano:1993yx,Coriano:1998ge,Leinweber:1995fn,Huang:1998wj,Harnett:2000fy,
Gubler:2010cf} is not necessary, once a dispersion relation is solved directly. 
This proposal, relying solely on analyticity of physical observables, has been applied 
to derivations of various nonperturbative quantities, covering hadronic properties of 
light resonances \cite{Li:2020ejs,Li:2021gsx,Zhao:2024drr,Mutuk:2024jvv,Li:2022qul,
Li:2024fko,Zeng:2026vyp} and heavy states \cite{Mutuk:2026rkq}, or in nuclear matter 
\cite{Mutuk:2025lak}. The new perspective has been implemented to constrain the hadronic 
vacuum-polarization contribution to the muon anomalous magnetic moment \cite{Li:2020fiz}, 
to accommodate neutral meson mixing parameters \cite{Li:2020xrz,Li:2022jxc} and to calculate 
$D$ and $B$ meson lifetimes \cite{Li:2023dqi}. Namely, the dispersive approach is applicable 
not only to observables defined by correlation functions, but to amplitudes with  
external hadronic states, as long as they are analytical.

We will generalize our formalism to studies of high-energy inclusive QCD processes, 
taking the proton structure functions in deeply inelastic scattering (DIS) as 
an example. The standard theoretical framework for these processes is the factorization 
theorem, which, however, breaks down at higher twists in most cases; nonfactorizable soft 
gluons resolve color structures of colliding or outgoing hadrons eventually, when stringent 
precision is demanded. It is difficult to gain rigorous estimates of the uncertainties from 
factorization-violating effects. The idea is similar to the dispersive analyses of the 
aforementioned heavy meson lifetimes \cite{Li:2023dqi}, which are also inclusive 
observables. It is based on analyticity of a double deeply virtual Compton scattering 
(DDVCS) amplitude on a complex baryon mass plane. The associated dispersion relation links 
the hadron- and quark-level descriptions for the same scattering amplitude evaluated at 
the masses $m_N$ of a heavy baryon and $m_Q\approx m_N$ of its heavy constituent quark, 
respectively. They correspond to the hadron and quark sides of QCD sum rules, and are
formulated in terms of dispersive integrals. The imaginary part of a forward DDVCS 
amplitude, which appears in the dispersive integral, is related to DIS cross section. 
For the quark-level description, OPE holds, which expresses the heavy quark DIS 
structure functions in orders of the strong coupling $\alpha_s$ and in powers of 
$1/Q$, $Q$ being the photon virtuality. The proton structure functions in the 
dispersive integral for the hadron-level description can then be obtained by solving the 
dispersion relation as an inverse problem with OPE inputs.

The OPE for the heavy quark DIS follows the 
short-distance factorization, so our approach differs conceptually from the factorization 
theorem established on the light cone. The inputs remain reliable even in the threshold 
region with the Bjorken variable $x_B\to 1$, because the effective Bjorken variable 
$x_N=m_px_B/m_N$ for the heavy baryon DIS (see the detail in Sec.~II), $m_p$ being the 
proton mass, is always below unity. There is thus sufficient phase space for real radiative 
corrections to fulfill infrared cancellation with virtual radiative corrections at leading 
power. Collinear divergences are absent, when both initial and final quarks are massive. 
Therefore, the threshold or endpoint region, where the heavy quark expansion for inclusive 
heavy hadron decays \cite{Chay:1990da} is usually jeopardized, does not cause a problem, 
and parton distribution functions (PDFs) need not be introduced. 
Though PDFs are not requested, they can be extracted from the solved structure functions 
as a byproduct, by treating the relevant factorization formulas as another inverse problem 
\cite{Hanninen:2026cea,Elmer:2026sbp}. In this respect, the application is similar to the 
determinations of PDFs and distribution amplitudes from the correlation functions with 
fictitious heavy quark fields in lattice QCD \cite{Detmold:2005gg,Detmold:2021qln}.

The OPE for the heavy quark DIS is prepared up to next-to-leading order (NLO) in 
$\alpha_s$ at leading power, with which we focus on valence quark contributions, and 
consider only the corresponding handbag diagram for the forward DDVCS. The setup 
matches the kinematic domain with the Bjorken variable $x_B$ being above, say, 0.2, 
and the photon virtuality $Q^2$ being above, say, 100 GeV$^2$. Contributions from gluonic 
partons in a proton are largely excluded for an intermediate $x_B$. Diagrams of other 
topologies for DDVCS, like a cat's ears diagram, are suppressed by powers of $1/Q$ relative 
to the dominant handbag diagram, since additional off-shell internal particles are involved 
in the former; hard gluon exchanges among valence quarks must be added into a 
cat's ears diagram to balance the initial and final quark momenta. It is easy to find that a 
cat's ears diagram for DDVCS, after a final-state cut is imposed, leads to multi-parton 
(higher-twist) contributions to a DIS. We will target the phase space enclosed by 
$x_B=0.18$-0.65 and $Q^2=10^2$-$10^4$ GeV$^2$, in which experimental data for the proton
structure functions are available for comparison.

Prior information is crucial for uniquely fixing solutions to an inverse problem 
\cite{Xiong:2022uwj}. Specific boundary conditions of an unknown force the emergence of 
a nontrivial solution distinct from the OPE input, although their dispersive integrals 
yield the roughly equal DDVCS amplitudes at a large mass. The nontrivial solution then 
captures nonperturbative dynamics in the low-mass region. It has been argued that the 
quark counting rules \cite{Brodsky:1973kr,Ball:2016spl} prescribe the endpoint behavior 
$(1-x_B)^{2n_s-1}=(1-x_B)^3$ of the structure function near the Bjorken variable $x_B\to 1$, 
where $n_s=2$ is the number of spectator partons in a proton. Note that the quark 
counting rules are not exact, and slight deviation has been observed in 
\cite{Ball:2016spl}. Another prior is motivated by the scaling, i.e., 
the mild dependence of the structure function on the photon virtuality $Q^2$. This feature 
is inherent in the perturbative (leading-power) piece in the OPE from the heavy quark 
regime, where it becomes insensitive to the variation of $Q^2$. It suggests that a solution 
to the dispersion relation can describe the structure function in a wide range of the baryon 
mass, from the nonperturbative region around the proton mass to a large-mass region.
At last, a physical solution for the structure function must obey positivity.

The proton mass $m_p$ is the only nonperturbative parameter at the leading-power 
precision, where we read the proton structure function $F_2$ off the corresponding 
solution to the dispersion relation. It is similar to the dispersive analysis of the
heavy meson lifetimes \cite{Li:2023dqi}, where we read the decay widths off the solutions
to the relevant dispersion relations at the heavy quark masses. It is surprising that 
our predictions for $F_2$ agree with the data in the considered broad $x_B$-$Q^2$ domain, 
and the agreement is prominent for the intermediate $x_B=0.25$ and 0.40. 
We then derive the proton structure function $F_L$, whose results are also consistent 
with the data with sizable experimental errors, and indicate the Callan-Gross 
relation $F_L\approx 0$ in the Bjorken limit. The same analysis with the 
pion mass as the nonperturbative parameter explains the data for the pion structure 
function $F_2^\pi$. The theoretical precision of our predictions can be enhanced 
systematically by including higher-power and higher-order contributions 
\cite{Zijlstra:1992qd} to the OPE. The inclusion of power corrections will bring 
in more nonperturbative parameters in terms of hadronic matrix elements of 
higher-dimensional local operators. Infrared renormalon effects, which generate additional 
power corrections \cite{Liao:2002cs}, can be taken into account too.

We point out that the structure functions acquired at large $x_B$ help pin down PDFs 
at high momentum fractions, where experimental constraints are rather loose compared to 
those at intermediate momentum fractions. As mentioned before, higher-twist multi-parton 
contributions to DIS can be deduced from the perturbative inputs for doubly heavy 
baryon DDVCS through cat's ears diagrams, from which corresponding multi-parton PDFs can 
be extracted. It is known that either computing these PDFs from first principles 
or fitting them from data is still at a preliminary stage \cite{Fedkevych:2025lgp}. 
Sticking to quark contributions at leading power, we expect that the neutron structure 
function $F_2^n$ is about $(2/3)^2+(1/3)^2+(1/3)^2=2/3$ times of the proton one under the 
spin-flavor $SU(6)$ symmetry, i.e., the ratio $F_2^n/F_2\approx 2/3$ for intermediate 
$x_B$ \cite{Ball:2016spl,Kuti:1971ph,Petratos:2000qp,Cocuzza:2026zoy}. The determination 
of $F_2^n/F_2$ from nucleus DIS data suffers a substantial model dependence 
\cite{Petratos:2000qp}. The $x_B$ dependence of the ratio can originate from 
subleading-power contributions to the OPE, which are associated with hadronic 
matrix elements depending on species of external baryons. The above extensions will highlight 
the potential to develop the dispersive method into a framework complimentary to the 
conventional factorization theorem.

The proton mass $m_p$ in the effective Bjorken variable $x_N=m_px_B/m_N$ for the heavy 
baryon DIS is about the soft scale associated with a heavy hadron. The 
value of $x_N$ would be so low for small $x_B$ that soft parton contributions become 
non-negligible. Both heavy and light constituents of a heavy baryon then
contribute, and the cross section can no longer be approximated by the heavy quark 
one owing to the soft parton contamination, rendering the dispersion relations  
invalid. The case differs from inclusive heavy hadron decays 
\cite{Chang:1998ze,Chang:1998ks}, where the heavy constituent maintains its role as an 
active parton in the whole kinematic region. Hence, our current setup is restricted to the 
region with finite $x_B$. A challenge for a dispersive analysis resides in the control 
of theoretical errors sourced from ill-posedness \cite{Zheng:2026zve}. This challenge could 
be overcome by available mature mathematical tools for attacking an inverse problem.   
Possible deviation from the quark counting rules i.e., from the power law 
$(1-x_B)^{2n_s-1}$ \cite{Ball:2016spl} also induces uncertainties. There are numerous 
subjects, including the improvement of the formalism and the survey of its phenomenological 
impacts, are worth investigations. Our endeavor represents the first 
attempt in this new direction.

The rest of the paper is organized as follows. In Sec.~II we elucidate how to derive 
the dispersion relation for a forward DDVCS amplitude. The resultant integral equation 
clearly connects the nonperturbative DIS structure functions to the perturbative inputs 
in the heavy quark regime. We recollect the inverse matrix method \cite{Li:2021gsx} for 
solving an ill-posed integral equation in Sec.~III. The priors are employed to settle 
the behaviors of a solution at low and high masses. The perturbative inputs 
up to NLO in $\alpha_s$ at leading power in $1/Q$ are presented in Sec.~IV. We discuss
the numerical outcomes of the proton structure functions $F_2$ and $F_L$ within the
ranges $x_B=0.18$-0.65 and $Q^2=10^2$-$10^4$ GeV$^2$ and their implications. The results for 
the pion structure function $F_2^\pi$ demonstrate the straightforward 
applications to DIS of various hadrons. Section V summarizes our observations 
and highlights the prospects of this innovative approach.

\section{DISPERSION RELATION}

We first construct the dispersion relation for a DDVCS amplitude based on its analyticity. 
We do not start with a correlation function in QCD sum rules 
\cite{Hsieh:2004ee,Hsieh:2009qj}, which is defined by sandwiching four currents (two baryon 
currents for the initial and final protons and two electromagnetic currents) with the vacuum 
states. This correlation function exhibits complicated branch cuts from the real protons and 
real intermediate states. It is also difficult to isolate proton contributions from those of 
excited states due to the dense baryon mass spectrum. 

The DIS structure functions $F_2$ and $F_L$ are related to the imaginary part 
of the amplitude for a forward DDVCS $\gamma^*(q)+ p(P) \to \gamma^*(q)+ p(P)$,
$q$ ($P$) being the momentum of the virtual photon $\gamma^*$ (the proton $p$) with the 
invariant $q^2=-Q^2$ ($P^2=m_p^2$). The function $F_{2}$ denotes the dominant scattering 
via a transversely polarized virtual photon, and $F_{L}$ acts as a direct measure 
of longitudinal photon absorption. The cross section for the unpolarized lepton-proton DIS 
$\ell(k) + p(P) \to \ell + X$ is then written, in terms of $F_2$ and $F_L$, as
\begin{eqnarray}
\frac{d^{2}\sigma }{dx_{B}dQ^{2}}=\frac{4\pi \alpha ^{2}}{x_{B}Q^{4}}
\left[\left(1-y+\frac{y^{2}}{2}-\frac{m_p^{2}x_{B}^{2}y^{2}}{Q^{2}}\right)F_{2}(x_{B},Q^{2})
-\frac{y^{2}}{2}F_{L}(x_{B},Q^{2})\right],\label{cs}
\end{eqnarray}
where $\alpha$ is the fine structure constant, and the inelasticity parameter 
$y=P \cdot q/P \cdot k$ denotes the fraction of energy lost by the incoming lepton $\ell$
with the momentum $k$. 

The Lorentz invariants from the proton kinematics contain $P^2=m_p^2$ and $P\cdot q$. 
Choosing the rest frame of the proton, we have $P\cdot q=m_pq^0$ with the photon energy $q^0$, 
such that the proton kinematics reduces to a single scale $m_p$. Given the external Bjorken 
variable $x_B$ and the photon virtuality $Q^2$, the photon energy is fixed to 
$q^0=Q^2/(2x_B m_p)$. We then consider the heavy baryon DIS with the photon kinematics 
specified above, which designates an effective Bjorken variable 
\begin{eqnarray}
x_N\equiv\frac{Q^2}{2P_N\cdot q}=\frac{m_p}{m_N}x_B,
\end{eqnarray}
$P_N=(m_N,{\bf 0})$ being the heavy baryon momentum in its rest frame. It is apparent that 
$x_N$ approaches $x_B$ for the proton DIS as $m_N\to m_p$. The advantage of the baryon
rest frame is then manifest; in other frames, like the lepton-baryon center-of-mass frame, 
$P_N$ carries more than one components ($P_N^0$ and $P_N^z$, for instance), so $P_N^2$ and 
$P_N\cdot q$ stand for different scales. The derivation of the dispersion relation for the 
DDVCS amplitude would be intricate. 

The theoretical setup allows us to work on the forward DDVCS amplitude $\Pi(m_N)$ defined 
on a complex baryon mass $m_N$ plane. Suppose that we compute $\Pi(m_N)$ at a sufficiently 
large baryon mass $m_N$. The active quark scattered by the virtual photon is assigned as 
the heavy constituent carrying the momentum $P_Q=(m_Q,{\bf 0})$ with the mass 
$m_Q\approx m_N$. The resultant fictitious "heavy proton" resembles a $\Lambda_c^+$ 
($\Sigma_b^+$), when a $u$ ($d$) quark is regarded as the heavy one. The OPE  for the 
forward DDVCS amplitude $\Pi(m_N)$ can be formulated systematically by means of the 
heavy quark dynamics, whose leading term corresponds to a perturbative contribution 
and the rest is suppressed by powers of $1/Q$. The amplitude for a heavy up (down) quark 
is roughly equal to the one for the heavy baryon multiplied by a charge factor 
$(2/3)^2=4/9$ ($(1/3)^2=1/9$), since the perturbative contributions from the heavy up and 
down quarks differ only by their electric charge factors. The sum $\Pi^{\rm p}(m_Q)$ of the 
perturbative amplitudes over the three valence quarks for a proton is then the same as 
$\Pi(m_N)$ for the heavy baryon up to power corrections, 
\begin{eqnarray}
\Pi(m_N)=\Pi^{\rm p}(m_Q)+O(1/Q^2).\label{app}
\end{eqnarray} 

Note that power corrections begin at $1/Q^2$ in the present case. These corrections are 
associated with hadronic matrix elements formed by local operators of various dimensions. 
For example, one of the dimension-5 matrix elements, $\langle N|\bar Q(iD_\perp)^2 Q|N\rangle$ 
with the transverse component $D_\perp$ of a covariant derivative, is attributed to the Fermi 
motion of the heavy quark $Q$ inside the heavy baryon. It can arise from 
the expansion of the internal heavy quark propagator between the two virtual photons in 
the forward DDVCS in small intrinsic transverse momenta $k_\perp$,
\begin{eqnarray}
\frac{1}{2P_Q\cdot q-Q^2-k_\perp^2+i\epsilon}&\approx&
\frac{1}{2P_Q\cdot q-Q^2+i\epsilon}
+\frac{k_\perp^2}{(2P_Q\cdot q-Q^2+i\epsilon)^2}\nonumber\\
&\propto&\frac{1}{1-Q^2/(2P_Q\cdot q)+i\epsilon}+
\frac{x_Bm_p}{m_Q}\frac{1}{[1-Q^2/(2P_Q\cdot q)+i\epsilon]^2}\frac{k_\perp^2}{Q^2}.
\label{kperp}
\end{eqnarray}
The imaginary part of the first term gives the leading perturbative contribution to
the heavy quark DIS,
\begin{eqnarray}
{\rm Im}\Pi^{{\rm p}(0)}(m_Q)=\delta\left(1-\frac{Q^2}{2m_Q q^0}\right).\label{p0}
\end{eqnarray}
The second term is down by $1/Q^2$ explicitly, and its imaginary part leads to the
first derivative of the $\delta$-function, $\delta'(1-Q^2/(2m_Q q^0))$.
Here we keep only the perturbative contribution $\Pi^{\rm p}(m_Q)$ in Eq.~(\ref{app}) as 
the input for the dispersion relation, because our goal is to illustrate a new 
framework for inclusive QCD processes. The impact of the power corrections will be 
explored in the future.

We express the analytical scattering amplitudes $\Pi(m_N)$ and $\Pi^{\rm p}(m_Q)$ in 
Eq.~(\ref{app}) as the contour integrals,  
\begin{eqnarray}
\frac{1}{2\pi i}\oint dm\frac{\Pi(m)}{m-m_N}=\Pi(m_N)
=\Pi^{\rm p}(m_Q)
=\frac{1}{2\pi i}\oint dm\frac{\Pi^{\rm p}(m)}{m-m_Q}.\label{di1}
\end{eqnarray}
The contour consists of two pieces of horizontal paths above and below the branch cut 
along the positive real axis on the complex mass plane, and a big circle $C$ of radius $R$ 
\cite{Li:2020ejs}. The contour integral on the left-hand (hadron) side of Eq.~(\ref{di1}) 
is decomposed into
\begin{eqnarray}
\frac{1}{2\pi i}\oint dm\frac{\Pi(m)}{m-m_N}=
\frac{1}{\pi}\int_{m_0}^R dm\frac{{\rm Im}\Pi(m)}{m-m_N} 
+\frac{1}{2\pi i}\int_C dm\frac{\Pi^{\rm p}(m)}{m-m_N},\label{di2}
\end{eqnarray}
in which the imaginary part ${\rm Im}\Pi(m)$, involving nonperturbative dynamics from the 
small $m$ region of the branch cut, is regarded as an unknown. The threshold 
$m_0=Q^2/(2q^0)$, i.e., the lower bound in the dispersive integral, sets the minimal baryon 
mass for producing a real final state in DIS, $(P_N+q)^2\ge m_N^2$. The integrand along 
the big circle $C$ has been replaced by $\Pi^{\rm p}(m)$, for the perturbative evaluation 
of $\Pi(m)$ is trustworthy for $m$ far away from the physical low-$m$ region. The 
same decomposition applies to the contour integral on right-hand side of Eq.~(\ref{di1}) 
with the same threshold $m_0$ owing to the condition $(P_Q+q)^2\ge m_Q^2$ at the quark level. 
The equality in Eq.~(\ref{di1}) then leads to the dispersion relation 
\begin{eqnarray}
\int_{m_0}^{R}\frac{{\rm Im}\Pi(m)}{m_N-m}dm
=\int_{m_0}^{R}\frac{{\rm Im}\Pi^{{\rm p}}(m)}{m_Q-m}dm,\label{di4}
\end{eqnarray}
where the contributions of $\Pi^{\rm p}(m)$ from the big circle $C$ on the two sides
have canceled.

The imaginary part ${\rm Im}\Pi^{{\rm p}}(m)$ of the perturbative piece is the input to the 
integral equation (\ref{di4}). We work on the proton structure function $F_2$, and the 
formulas for another structure function $F_L$ are similar. The corresponding leading-order 
result has been given in Eq.~(\ref{p0}), which stems from the on-shell condition of the 
scattered quark $\delta((P_Q+q)^2-m_Q^2)$. The NLO contribution consists of the virtual and 
real pieces, 
\begin{eqnarray}
{\rm Im}\Pi^{{\rm p}(1)}(m_Q)={\rm Im}\Pi_v^{{\rm p}(1)}(m_Q)+{\rm Im}\Pi_r^{{\rm p}(1)}(m_Q),
\label{one}
\end{eqnarray}
respectively, with
\begin{eqnarray}
{\rm Im}\Pi_v^{{\rm p}(1)}(m_Q)&=&\frac{\alpha_s(\mu^2)}{2\pi}C_F
[S_2(m_Q)+V_2(m_Q)]\delta\left(1-\frac{Q^2}{2m_Q q^0}\right)\nonumber\\
{\rm Im}\Pi_r^{{\rm p}(1)}(m_Q)&=&\frac{\alpha_s(\mu^2)}{2\pi}C_F
\frac{\hat s(m_Q)-m_Q^2}{8\hat s(m_Q)}\frac{f_2^Q(m_Q)}{N_2(m_Q)}.\label{piv}
\end{eqnarray}
The quark-level invariant $\hat s$ is defined as $\hat s=(P_Q+q)^2=m_Q^2+2m_Qq^0-Q^2$, 
and the expressions of $S_2$, $V_2$, $N_2$ and $f_2^Q$ can be found in the
Appendices of \cite{Kretzer:1998ju}. The renormalization scale in $\alpha_s$ will be 
chosen as the hard scale $\mu=Q$ involved in the DIS. As claimed before, no collinear 
divergences are associated with the initial and final heavy quarks, and the soft divergences 
cancel between the real and virtual corrections, such that Eq.~(\ref{one}) is infrared safe.

It is noticed that the function ${\rm Im}\Pi_r^{{\rm p}(1)}(m)$ increases with the mass $m$, 
so an ultraviolet subtraction is implemented; we introduce a subtracted unknown
\begin{eqnarray}
\Delta\rho(m,\Lambda)=\rho(m)
-{\rm Im}\Pi^{\rm p}(m)\{1-\exp[-(m-m_0)^\gamma/\Lambda^\gamma]\},
\label{sub}
\end{eqnarray}
with $\rho(m)\equiv{\rm Im}\Pi(m)$ and the power $\gamma > 0$. The function 
$1-\exp[-(m-m_0)^\gamma/\Lambda^\gamma]$ in the subtraction term approaches unity as  
$m\gg\Lambda$ to guarantee the diminishing of $\Delta\rho(m,\Lambda)$ at large $m$. 
The arbitrary scale $\Lambda$ characterizes how fast ${\rm Im}\Pi(m)$ transits to the 
perturbative ${\rm Im}\Pi^{\rm p}(m)$. Certainly, a physical solution ought to 
be insensitive to the variation of $\Lambda$ \cite{Li:2021gsx} as shown in Sec.~IV. 
The radius $R$ in Eq.~(\ref{di4}) can be pushed toward the infinity, when the dispersion
relation is formulated in terms of the subtracted unknown,
\begin{eqnarray}
\int_{m_0}^\infty dm\frac{\Delta\rho(m,\Lambda)}{m_N-m}
&=&\int_{m_0}^\infty dm \frac{{\rm Im}\Pi^{{\rm p}}(m) 
\exp[-(m-m_0)^\gamma/\Lambda^\gamma]}{m_Q-m},\label{r20}
\end{eqnarray}
in which the dependence on the arbitrary radius $R$ has migrated to that on $\Lambda$.

The variable changes $m_N-m_0\approx m_Q-m_0=x\Lambda$ and $m-m_0=y\Lambda$ recast 
Eq.~(\ref{r20}) into
\begin{eqnarray}
& &\int_{0}^\infty dy\frac{\Delta\rho(y)}{x-y}
=\int_{0}^\infty dy \frac{{\rm Im}\Pi^{{\rm pert}}(y\Lambda+m_0) \exp(-y^\gamma)}{x-y}.
\label{r21}
\end{eqnarray}
The above equation aims at solving for a structure function, which depends on the 
dimensionless variable $y=(m-m_0)/\Lambda$. Equation~(\ref{r21}) is classified as the 
first kind of Fredholm integral equations, which has the typical form 
\begin{eqnarray}
\int_{0}^\infty dy\frac{\Delta\rho(y)}{x-y}= \omega(x),\label{su1}
\end{eqnarray}
with the unknown $\Delta\rho(y)$ and the input $\omega(x)$.

\section{Inverse Matrix Method}

We explicate the inverse matrix method \cite{Li:2021gsx} for solving the integral equation 
(\ref{su1}), whose solid mathematical ground is referred to \cite{CWG}. When the subtracted 
unknown $\Delta\rho(y)$ descends quickly enough with $y$ after the ultraviolet subtraction, 
the major contribution to the integral on the left-hand side comes from a finite range of 
$y$. It is then legitimate to expand the integral into a series in $1/x$ up to some power 
$N$ for a sufficiently large $|x|$ by inserting
\begin{eqnarray}
\frac{1}{x-y}=\sum_{j=1}^N \frac{y^{j-1}}{x^j}\label{ep1}
\end{eqnarray}
into Eq.~(\ref{su1}). The input $\omega(x)$ can also be expanded into a power series in 
$1/x$ for a large $|x|$, 
\begin{eqnarray}
\omega(x)=\sum_{i=1}^N \frac{b_i}{x^i},\label{bb1}
\end{eqnarray}
in the presence of the exponential factor $\exp(-y^\gamma)$ on the right-hand side 
of Eq.~(\ref{r21}).

We expand the subtracted unknown into
\begin{eqnarray}
\Delta\rho(y)=\sum_{i=1}^N a_iy^\alpha e^{-y}L_{i-1}^{(\alpha)}(y),\label{r1}
\end{eqnarray}
using a set of generalized Laguerre functions $L_i^{(\alpha)}(y)$ up to degree $N-1$,
which satisfies the orthogonality
\begin{eqnarray}
\int_0^\infty y^\alpha e^{-y}L_i^{(\alpha)}(y)L_j^{(\alpha)}(y)dy
=\frac{\Gamma(i+\alpha+1)}{i!}\delta_{ij}.
\label{or1}
\end{eqnarray}  
The maximal integer $N$ will be prescribed later, and the value of the index $\alpha$ depends
on the behavior of $\Delta\rho(y)$ near the threshold $y\to 0$. 
Substituting Eqs.~(\ref{ep1}), (\ref{bb1}) and (\ref{r1}) into Eq.~(\ref{su1}), and 
equating the coefficients of $1/x^i$ (remember that $|x|$ is still arbitrary despite being 
large), we arrive at the matrix equation $Ma=b$ with the matrix elements
\begin{eqnarray}
M_{ij}=\int_{0}^\infty dy y^{i-1+\alpha}e^{-y}L_{j-1}^{(\alpha)}(y),\label{m2}
\end{eqnarray}
the unknown vector $a=(a_1, a_2,\cdots,a_N)$ and the input vector $b=(b_1,b_2,\cdots,b_N)$, 
where $i$ and $j$ run from 1 to $N$.

If the inverse matrix $M^{-1}$ exists, one can get a solution for $a$ from the input $b$ 
via $a=M^{-1}b$ straightforwardly, and construct the subtracted unknown in Eq.~(\ref{r1}). 
In principle, the true solution is approached by increasing the number of polynomials $N$ in 
Eq.~(\ref{r1}). Nevertheless, both $i$ and $j$ have to stop at a finite $N$ in a practical 
application, for the determinant of $M$ decreases with its dimension. The series $a_i$ 
would grow eventually with $i$ owing to the amplification by the huge elements of $M^{-1}$, 
though the series $b_i$ decreases. The convergence of $a_i$ is then lost, and a
solution deviates from the true one. This is a generic feature of an ill-posed
inverse problem. We work on a matrix $M$ with its dimension not greater than 
$20\times 20$ to avoid the divergence of $M^{-1}$. The optimal choice of $N$ is  
set to the integer labelling the minimal $a_N$, beyond which the series $a_i$ 
starts to increase. Fortunately, most $N$'s for the best convergence turn out to
be located in the range $N=15$-18 in the considered $x_B$-$Q^2$ phase space. We have 
confirmed that the positivity of a solved structure function is always met.

It is known that priors help determine solutions to an inverse problem. First, the low 
end of the mass $m\approx m_p$, where we will read the proton structure function off the 
solution of $\rho(m)$, corresponds to the threshold region $x_B\to 1$; for $m\approx m_p$, 
we have the relation $m-m_0\approx m_p-m_0=(1-x_B)m_p$. The quark counting rules 
\cite{Brodsky:1973kr,Ball:2016spl} dictate the threshold behavior of a structure function 
$\sim (1-x_B)^3$, which is equivalent to 
$\rho(m)\sim (m-m_0)^3$ for small $m$. We must require $\gamma\ge 3$, such that the 
function $1-\exp[-(m-m_0)^\gamma/\Lambda^\gamma]$ decreases faster than $(m-m_0)^3$ at 
small $m$, and $\Delta\rho(m,\Lambda)$ retains the threshold behavior of the unknown 
$\rho(m)$. The power $\gamma$ will be set to $\gamma=3$ below. We have checked that 
subtractions with powers higher than 3, like $\gamma=4$, alter our predictions by only 
few percent (less than 5\%). Note that a generalized Laguerre polynomial diminishes as 
$L_{i-1}^{(\alpha)}(y)\sim y^\alpha$ at small $y$, implying $\Delta\rho(y)\sim y^{2\alpha}$. 
Therefore, we designate $\alpha=3/2$ according to the aforementioned quark counting rules 
$\rho(y)\propto y^3$ near the threshold.

Second, a solution of $\rho(m)$ is supposed to describe a structure function for 
an arbitrary baryon mass, from the nonperturbative low-mass region with $m\approx m_p$ to  
a heavy-mass region. However, the mass up to which it is applicable cannot be too high. 
Otherwise, a solution is forced to encode too much information on DIS of various heavy 
baryons that does not match the finite precision of the OPE input truncated at leading 
power and NLO, thus ceasing to exist. Another prior that a structure function 
approximately scales in the photon virtuality $Q^2$ sets the order of magnitude of this
upper bound in $m$. We observe that the scales involved in the dimensionless perturbative 
input in Eq.~(\ref{piv}) can be organized into ratios, like 
$Q^2/(P\cdot q)\sim Q^2/(m q^0)$ and $P^2/(P\cdot q)\sim m/q^0$. The former scales like a 
constant in the case of light baryon DIS (for small $m$) and the latter is negligible, 
ensuring a moderate $Q^2$ dependence. When $m$ increases, the ratio $m/q^0=2x_Bmm_p/Q^2$ 
may cause a $Q^2$ dependence. To have a constant ratio $m/q^0$, the large $m$ we can 
reach should scale like $q^0\sim Q^2/m_p$. Then $Q^2/(m q^0)\sim O(m_p^2/Q^2)$ becomes 
negligible compared to $m/q^0\sim O(1)$, and the $Q^2$ independence is maintained. In 
this way the scaling behavior of a structure functions persists for both small and large 
baryon masses up to $m\sim q^0$. For the numerical analysis in the next section, we 
choose the mass bound as $m_b\approx 2x_BQ^2/m_p$, such that the effective Bjorken variable 
$x_Bm_p/m=(m_p/Q)^2/2$ is not lower than half of the soft parameter involved in the heavy 
baryon DIS to exclude soft parton contributions. Note that this bound should be understood
as an order-of-magnitude estimate, so theoretical uncertainties associated with its 
variation need to be taken into account. We will solve the integral 
equation (\ref{r21}) with the above two pieces of prior information.

\section{DIS STRUCTURE FUNCTIONS}

\subsection{Proton Structure Function $F_2$}

We provide the input coefficients $b_i$ for the proton structure 
function $F_2$. The virtual corrections proportional to the $\delta$-function give, up to NLO, 
\begin{eqnarray}
\left[1+\frac{\alpha_s(Q^2)}{2\pi}C_F\left(S_2+V_2\right)\right]
\frac{Q^2}{2q^0}\frac{1}{m_Q-Q^2/(2q^0)}
&=&\left[1+\frac{\alpha_s(Q^2)}{2\pi}C_F\left(S_2+V_2\right)\right]
\frac{Q^2}{2\Lambda q^0}\frac{1}{x},\label{del}
\end{eqnarray}
where the variable change $m_Q=x\Lambda+m_0$ has been inserted. It indicates that the 
virtual corrections contribute only to the input coefficient of the first power in $1/x$. 
The input from the NLO real corrections is expanded, by means of Eq.~(\ref{ep1}), as
\begin{eqnarray}
\frac{\alpha_s(Q^2)}{2\pi}C_F\sum_{i=1}^N\frac{1}{x^i}
\int_{0}^\infty dy y^{i-1}{\rm Im}\Pi_r^{{\rm p}(1)}(y\Lambda+m_0) \exp(-y^3).
\end{eqnarray}

The input vector $b$ assembles the components
\begin{eqnarray}
b_1&=&\left[1+\frac{\alpha_s(Q^2)}{2\pi}C_F\left(S_2+V_2\right)\right]
\frac{Q^2}{2\Lambda q^0}+\frac{\alpha_s(Q^2)}{2\pi}C_F
\int_{0}^\infty dy {\rm Im}\Pi_r^{{\rm p}(1)}(y\Lambda+m_0) \exp(-y^3),\nonumber\\
b_{i>1}&=&\frac{\alpha_s(Q^2)}{2\pi}C_F
\int_{0}^\infty dy y^{i-1}{\rm Im}\Pi_r^{{\rm p}(1)}(y\Lambda+m_0)\exp(-y^3).\label{ib}
\end{eqnarray}
The above expressions evince how the singular virtual contributions (proportional to a 
$\delta$-function) are smoothed into finite results 
through the convolution with the kernel $1/(x-y)$. This kernel plays a role similar to 
a PDF in the factorization theorem. The one-loop running coupling constant $\alpha_s(Q^2)$
is employed with the number of active quarks $n_f=5$ and the corresponding QCD scale
$\Lambda_{\rm QCD}=0.21$ GeV \cite{Zhong:2021epq}. We then calculate the matrix elements 
$M_{ij}$ in Eq.~(\ref{m2}), solve for the unknown vector $a$ from $a=M^{-1}b$, and 
build a solution in Eq.~(\ref{r1}) as an expansion in generalized Laguerre polynomials. 
Note again that the integer $N$ for the polynomial expansion indexes the minimal 
element $|a_N|$ in the vector $a$. The input coefficients $b_i$ depend on the external 
variables $x_B$ and $Q^2$, so do the solved coefficients $a_i$.

A solution from Eq.~(\ref{r21}) depends on the arbitrary scale $\Lambda$ obviously.
An important ingredient of the dispersive approach is to search for a stability window, in 
which a solution is relatively insensitive to the variation of $\Lambda$ \cite{Li:2022jxc}. 
In other words, the stability identifies a physical solution. It has been argued in the previous 
section that a physical solution holds up to the mass $m_b\approx 2x_BQ^2/m_p$. A $\pm 20\%$ 
variation around $m_b$ is postulated to allow room for theoretical uncertainties. We 
thus manage to get a stability window in $\Lambda$ for $\rho(m)$ evaluated for a set 
$I$ of five $m$ values, which pick up $0.8m_b$, $0.9m_b$, $\cdots$, and $1.2m_b$. Extending 
the set $I$ to cover a wider interval of $m$, say, between $0.7m_b$ and $1.3m_b$, changes the 
outcomes for the proton structure function $F_2$ by less than 5\%. Explicitly, 
we search for a local minimum of the difference between the two solutions corresponding to 
$\Lambda+\Delta\Lambda$ and $\Lambda$ with a fixed step $\Delta\Lambda$,
\begin{eqnarray}
d=\Sigma_i|\rho_{\Lambda+\Delta\Lambda}(m_i)-\rho_\Lambda(m_i)|_{m_i\in I},\label{dev}
\end{eqnarray}
by varying $\Lambda$.

It will be demonstrated that a local minimum of $d$ in Eq.~(\ref{dev}) indeed appears at  
some value of $\Lambda=\Lambda_m$. The series $a_i$ corresponding to $\Lambda=\Lambda_m$ 
is selected to construct the physical solution of $\rho(m)$, which gives rise to the 
central value of our predictions for $F_2$. Solutions corresponding to $\Lambda$ in the 
neighborhood of the local minimum will be pinched, namely, close to each other around the 
mass $m\approx  m_b$. They are also restricted to zero by the threshold behavior 
$\sim(m-m_0)^3$  at small $m\approx m_0$ (recall the prior information on a 
structure function from the quark counting rules). As a consequence, the solutions 
for $\Lambda$ near $\Lambda_m$ are all similar, describing the structure function for an 
arbitrary baryon mass from the nonperturbative region with $m\approx m_p$ to the region 
with $m\sim q^0$. Results of $F_2$ read off these solutions are roughly $Q^2$-independent, 
and constitute the range of our predictions. Increasing the scale $\Lambda$ more 
reduces the difference $d$, which, however, does not lead to physical solutions. It has been 
elaborated \cite{Li:2021gsx} that as $\Lambda$ becomes large enough, the $\Lambda$ dependence 
disappears with the diminishing of perturbative inputs. This can be seen from, for instance, 
the $1/\Lambda$ suppression factor in Eq.~(\ref{del}) for the virtual contributions. 
Hence, the absolute minimum of the difference $d$ leads the trivial null solution, 
which is not physical.

After acquiring the physical solution for $\rho(m)$, we simply read off the proton structure 
function $F_2(x_B,Q^2)$ at the mass $m=m_p$. Rewriting the argument $y$ as
\begin{eqnarray}
y=\frac{m-m_0}{\Lambda_m}=\frac{m_p-Q^2/(2q^0)}{\Lambda_m}
=(1-x_B)\frac{m_p}{\Lambda_m},
\end{eqnarray}
we arrive at 
\begin{eqnarray}
F_2(x_B,Q^2)=\rho\left((1-x_B)\frac{m_p}{\Lambda_m}\right).
\end{eqnarray}
Since $\rho(m)$ describes the baryon DIS for an arbitrary mass,
we can also read off the structure functions for other baryons in principle, such as 
$\Lambda_c^+$, 
\begin{eqnarray}
F_2^{\Lambda_c^+}(x_B,Q^2)=\rho\left((1-x_B)\frac{m_{\Lambda_c^+}}{\Lambda_m}\right).
\end{eqnarray}
It is certainly difficult to measure this quantity experimentally. It may be 
more practical to consider the $\Lambda$ baryon DIS, whose data could be deduced via 
Sullivan processes.

We perform a case study with the Bjorken variable $x_B=0.25$ and the photon virtuality 
$Q^2=10^2$ GeV$^2$. Computing the perturbative inputs $b_i$ in Eq.~(\ref{ib}) for a typical 
$\Lambda$ value, say, $\Lambda=200$ GeV, we have the solution of $a=M^{-1}b$ from the 
inverse matrix method,
\begin{eqnarray}
a&=&(a_1,a_2,a_3,\cdots, a_{14},a_{15},a_{16},a_{17},a_{18})\nonumber\\
&=&(92.51, 66.80, 47.39, 32.91, \cdots, -0.4288, \
-0.2900, -0.09145, 0.1272, 0.3395),\label{sa}
\end{eqnarray} 
where the elements up to $a_{18}$ are listed. The series $a_i$ decreases with $i$, 
turns negative at $i=12$, and then grows monotonically. The last minimal magnitude of $a_i$ 
close to the high end of $N$ occurs at $i=16$, which sets the integer $N=16$ for the 
polynomial expansion. The series continues to increase with $i>18$, and goes out of control 
owing to ill-posedness.

\begin{figure}
\begin{center}
\includegraphics[scale=0.35]{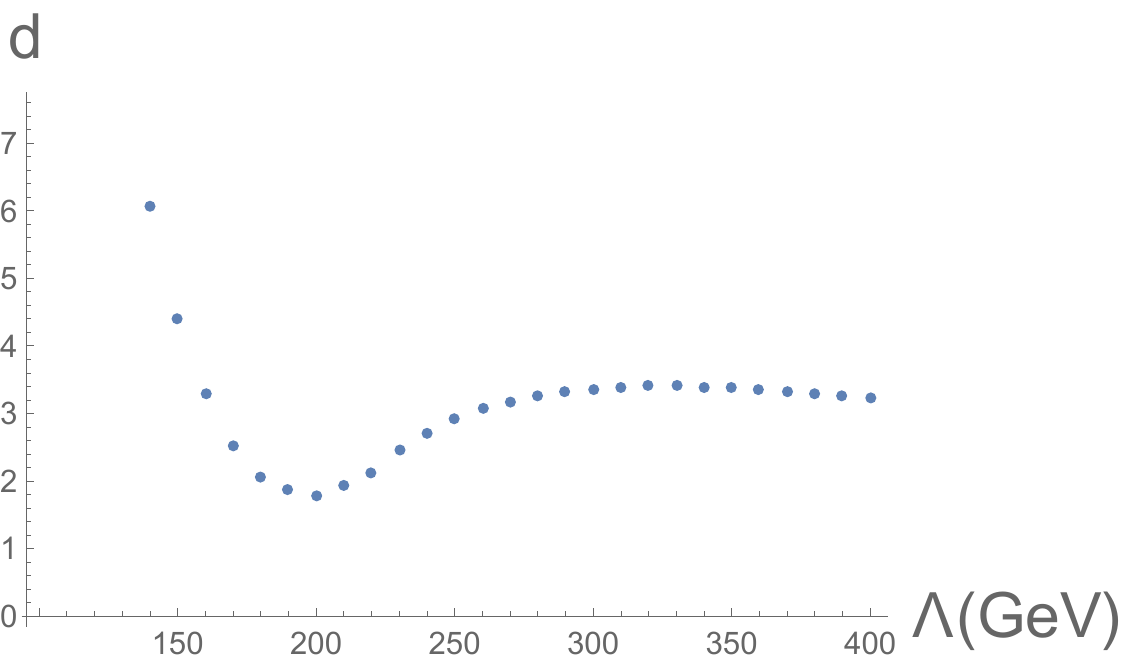}\hspace{1.0cm}
\includegraphics[scale=0.35]{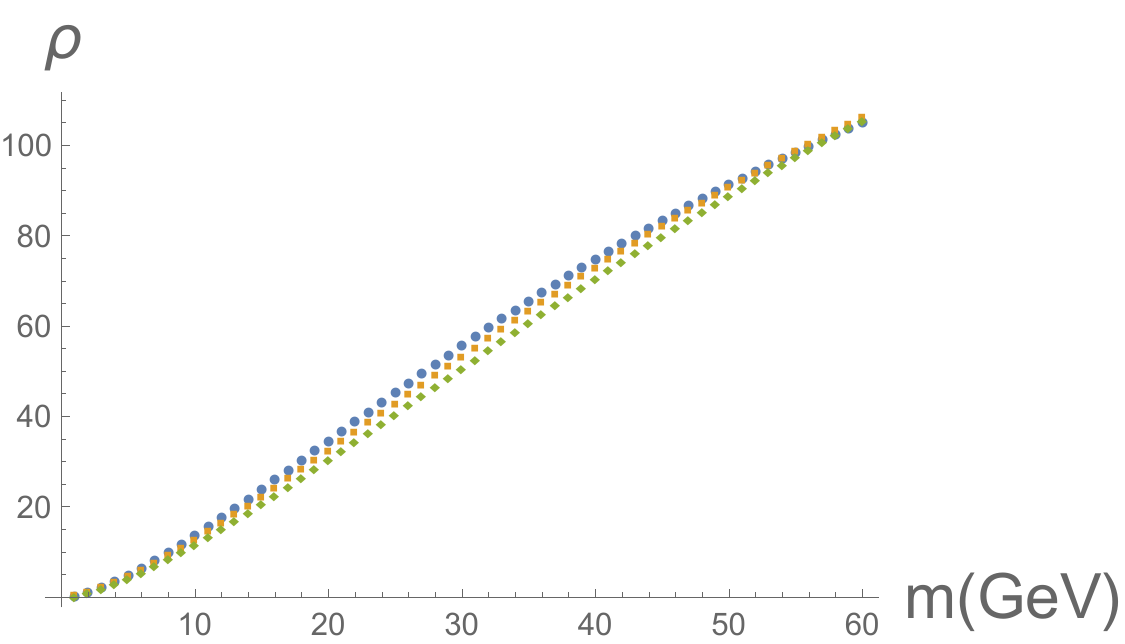}

(a) \hspace{7.0 cm} (b)
\caption{\label{fig1}
(a) $\Lambda$ dependence of the difference $d$ in Eq.~(\ref{dev}) for $x_B=0.25$ and 
$Q^2=10^2$ GeV. (b) $m$ dependencies of $\rho(m)$ for $x_B=0.25$, $Q^2=100$ GeV, and 
$\Lambda=170$ GeV ($N=16$, dots), 200 GeV ($N=16$, squares), 230 GeV ($N=17$, diamonds).}
\end{center}
\end{figure}

We then examine the difference $d$ in Eq.~(\ref{dev}) for the bound 
$m_b\approx 2x_BQ^2/m_p\approx 53$ GeV and the step $\Delta\Lambda=10$ GeV. The $\Lambda$ 
dependence of $d$ displayed in Fig.~\ref{fig1}(a) reveals a minimum at $\Lambda=200$ GeV, 
explaining why the solution of $a$ for $\Lambda=200$ GeV was presented in Eq.~(\ref{sa}). 
It implies that the solutions labeled by different $\Lambda$ around $\Lambda_m=200$ GeV 
are stable, i.e., similar to each other. The series $a$ in Eq.~(\ref{sa}) up to $N=16$ 
generates the solution $\rho(m)$ that leads to the central value of our predictions for 
$F_2$. To assess theoretical errors, we select the $\Lambda$ values, where $d$ is equal to 
its local maximum $d=3.3$ located at $\Lambda=320$ GeV. That is, the interval 
$\Lambda=160$-320 GeV makes up the span of our predictions. Without the aforementioned 
$\pm 20\%$ variation around $m_b$, this local minimum is narrower, and the resultant 
theoretical uncertainties are smaller, representing a more aggressive error estimation. The 
solutions of $\rho(m)$ for $\Lambda=170$, 200, and 230 GeV as functions of $m$ are exhibited 
in Fig.~\ref{fig1}(b), where the first and third $\Lambda$ values correspond to $d$ halfway 
between its local minimum and local maximum. It is observed that the three curves are 
close to each other. We read off the proton structure function $F_2$ defined at 
$m=m_p=0.938$ GeV \cite{PDG}, 
\begin{eqnarray}
F_2(x_B=0.25,Q^2=10^2\;{\rm GeV}^2)=0.27^{+0.05}_{-0.07},
%0.27^{+0.03}_{-0.02},
\end{eqnarray}
where the central (upper bound, lower bound) corresponds to $\Lambda=200$ GeV and $N=16$
($\Lambda=160$ GeV and $N=16$, $\Lambda=320$ GeV and $N=17$).

\begin{figure}
\begin{center}
\includegraphics[scale=0.4]{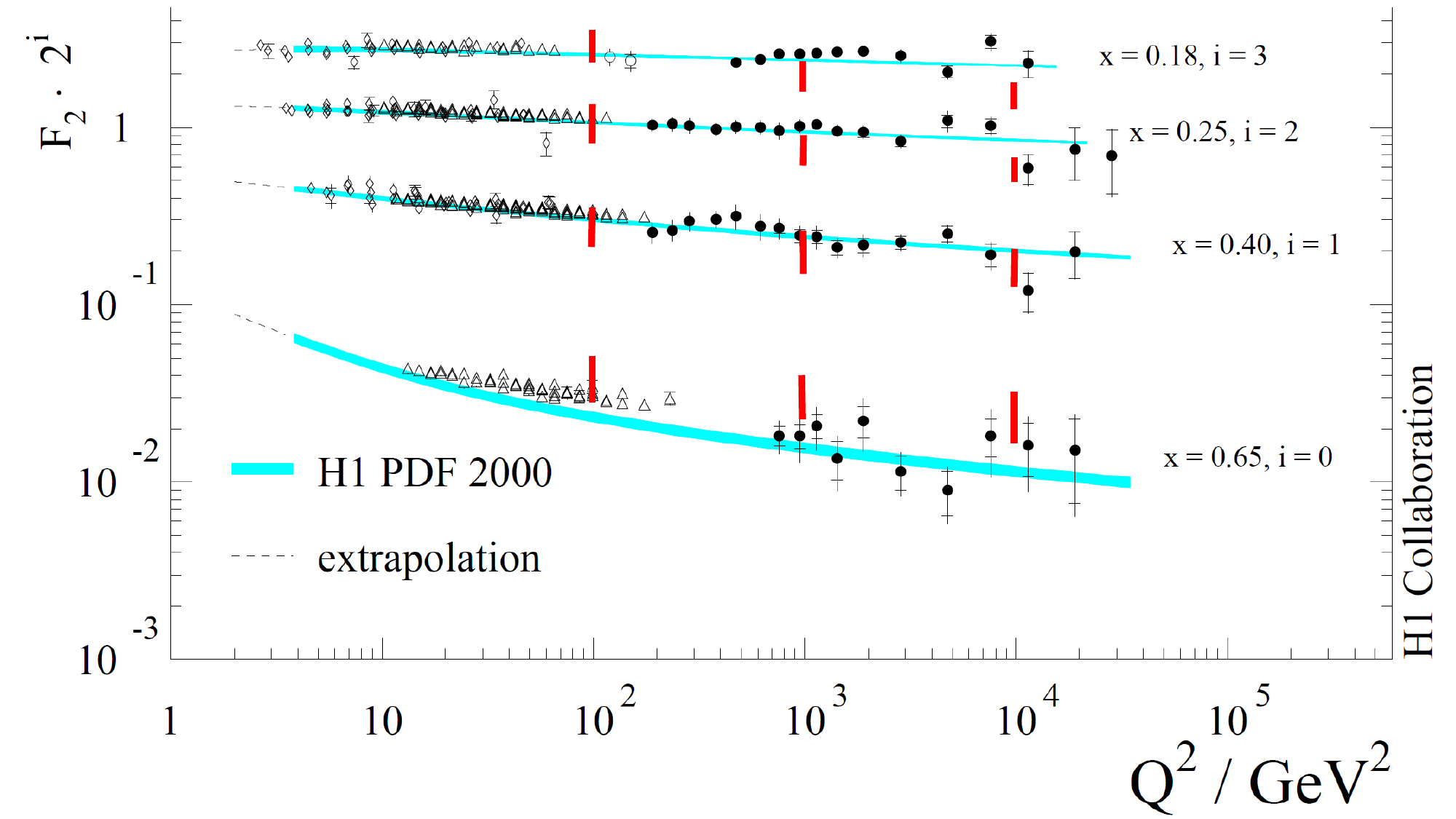}

\caption{\label{fig2}
Comparison of our predictions (red vertical bars) for the proton structure function 
$F_2(x_B,Q^2)$ with the data, which, together with the H1 PDF 2000 fits based on the 
factorization theorem, are quoted from \cite{H1:2003xoe}. Symbols for 
the various data sets are explained in \cite{H1:2003xoe}.}
\end{center}
\end{figure}

Repeating the procedure, we obtain the proton structure functions 
\begin{eqnarray}
F_2(x_B=0.25,Q^2=10^3\;{\rm GeV}^2)=0.19^{+0.02}_{-0.04},
%0.19^{+0.02}_{-0.01},
\end{eqnarray}
where the central value (upper bound, lower bound) corresponds to $\Lambda=2000$ GeV
($\Lambda=1600$ GeV, $\Lambda=3200$ GeV) and $N=18$, and
%where the central value (upper bound, lower bound) corresponds to $\Lambda=2000$ GeV
%($\Lambda=1600$ GeV, $\Lambda=2300$ GeV) and $N=18$, and
\begin{eqnarray}
F_2(x_B=0.25,Q^2=10^4\;{\rm GeV}^2)=0.15^{+0.02}_{-0.03},
%0.15^{+0.02}_{-0.01},
\end{eqnarray}
where the central value (upper bound, lower bound) corresponds to $\Lambda=19000$ GeV
($\Lambda=15000$ GeV, $\Lambda=32000$ GeV) and $N=18$. We also get $F_2$ for $x_B=0.40$ 
and $Q^2=10^2$, $10^3$ and $10^4$ GeV$^2$, and gather the outcomes in Table~\ref{tb}. 
The central values (upper bounds, lower bounds) correspond to 
$\Lambda=320$ GeV and $N=15$ ($\Lambda=240$ GeV and $N=15$,
$\Lambda=580$ GeV and $N=16$) for $Q^2=10^2$ GeV$^2$, to $\Lambda=3300$ GeV and $N=17$ 
($\Lambda=2400$ GeV and $N=17$, $\Lambda=6000$ GeV and $N=16$) for $Q^2=10^3$ GeV$^2$, and 
to $\Lambda=32000$ GeV and $N=17$ ($\Lambda=24000$ GeV and $N=17$, $\Lambda=60000$ GeV and 
$N=16$) for $Q^2=10^4$ GeV$^2$. 
It is found that the predicted $F_2$ in the above $x_B$-$Q^2$ 
domain agree well with the data \cite{H1:2003xoe} as shown in Fig.~\ref{fig2}.

\begin{table}
\begin{center}
 \begin{tabular}{c|c |c| c}
 \hline
  $F_2(x_B,Q^2)$ & $Q^2=10^2$ GeV$^2$  & $Q^2=10^3$ GeV$^2$ & $Q^2=10^4$ GeV$^2$ \\
  \hline
  $x_B=0.18$ & $0.40^{+0.06}_{-0.11}$ & $0.26^{+0.03}_{-0.06}$ & $0.20^{+0.02}_{-0.04}$ \\
  \hline
  $x_B=0.25$ & $0.27^{+0.05}_{-0.07}$ & $0.19^{+0.02}_{-0.04}$ & $0.15^{+0.02}_{-0.03}$ \\
  \hline
  $x_B=0.40$ & $0.14^{+0.03}_{-0.04}$ & $0.10^{+0.02}_{-0.03}$ & $0.086^{+0.013}_{-0.027}$ \\
  \hline
  $x_B=0.65$ & $0.039^{+0.012}_{-0.011}$ 
  & $0.032^{+0.007}_{-0.010}$ & $0.026^{+0.006}_{-0.008}$ \\
  \hline
 \end{tabular}
\caption{Predictions for the proton structure function $F_2(x_B, Q^2)$.}
\label{tb}
\end{center}
\end{table}

We investigate the structure function $F_2$ at a higher Bjorken variable
$x_B=0.65$ and collect the results in Table~\ref{tb}. The central values (upper bounds, 
lower bounds) correspond to $\Lambda=560$ GeV and $N=18$ ($\Lambda=400$ GeV and $N=16$, 
$\Lambda=100$ GeV and $N=20$) for $Q^2=10^2$ GeV$^2$, to $\Lambda=5600$ GeV and $N=18$ 
($\Lambda=4000$ GeV and $N=16$, $\Lambda=10^4$ GeV and $N=20$) for $Q^2=10^3$ GeV$^2$, and 
to $\Lambda=56000$ GeV and $N=18$ ($\Lambda=40000$ GeV and $N=16$, $\Lambda=10^5$ GeV and 
$N=20$) for $Q^2=10^4$ GeV$^2$. 
The theoretical errors are more significant, because the local minima of the difference $d$ 
are shallower and broader than in the previous cases. The integer $N=20$ for the lower bounds 
means that these values are less reliable. Our predictions are slightly above the data, but 
the consistency is still satisfactory in view of the sizable experimental uncertainties. 
Equation~(\ref{kperp}) indicates that power corrections to the OPE inputs are proportional 
to $x_B$. We thus speculate that power-suppressed terms have become relevant at $x_B=0.65$, 
and need to be included to improve the consistency.

At last, we attempt the predictions for the proton structure function $F_2$ at a smaller Bjorken 
variable $x_B=0.18$, which are listed in Table~\ref{tb}. The central values 
(upper bounds, lower bounds) correspond to $\Lambda=140$ GeV and $N=16$ ($\Lambda=120$ GeV 
and $N=15$, $\Lambda=220$ GeV and $N=17$) for $Q^2=10^2$ GeV$^2$, to $\Lambda=1400$ GeV
($\Lambda=1100$ GeV, $\Lambda=2200$ GeV) and $N=18$ for $Q^2=10^3$ GeV$^2$, and to 
$\Lambda=14000$ GeV ($\Lambda=11000$ GeV, $\Lambda=22000$ GeV) and $N=18$ for $Q^2=10^4$ 
GeV$^2$. The above results disclose a steeper slope in $Q^2$ than the data do, suggesting that 
gluonic partons begin to play a role at this lower Bjorken variable. It is likely that 
the gluon splitting into a quark-antiquark pair $g\to q\bar q$ \cite{Laenen:1992zk} 
may provide additional 
contributions to alleviate the $Q^2$ dependence. It is observed that all the above $\Lambda$ 
values from the stability windows scale more or less with $Q^2$. Namely, for a 
variation of $Q^2\to \beta Q^2$, the stability window in $\Lambda$ is simply shifted by 
the multiplication factor $\beta$. It is due to the feature that the perturbative input mainly 
depends on the ratio $m/q^0\propto \Lambda/Q^2$. We emphasize that only two pieces of prior 
information are incorporated in the derivation, 
so it is nontrivial to accommodate the data in the wide $x_B=0.18$-0.65 and 
$Q^2=10^2$-$10^4$ GeV$^2$ ranges.

\subsection{Proton Structure Function $F_L$}

We analyze another proton structure function $F_L$ in the dispersive formalism, which 
is related to the structure function $F_1$ in an alternative decomposition of DIS cross 
section via $F_L=F_2-2x_BF_1$. The Callan-Gross relation $F_2\approx 2x_BF_1$, which is 
supposed to hold for a spin-$1/2$ quark being scattered in the Bjorken limit, i.e., at 
a finite Bjorken variable $x_B$ and a high photon virtuality $Q^{2}$, implies the 
smallness of $F_L$. Deviation from the Callan-Gross relation is expected at low $x_B$, 
where gluonic partons dominate the scattering. Precise measurements of $F_L$ come from the 
H1 and ZEUS collaborations, which are, however, concentrated in the small $x_B$ region 
($x_B < 0.01$) \cite{H1:2008rkk}. Most intermediate $x_B$ data from fixed targets sit at 
lower $Q^2$ ($Q^2 < 20 \text{GeV}^2$) \cite{Tvaskis:2016uxm}. The future Electron-Ion 
Collider (EIC) is specifically designed to implement Rosenbluth separations across a broad 
center-of-mass energy configuration, offering unprecedented precision for $F_{L}$ in the 
intermediate $x_B$ and high $Q^2$ region. Here we will compare our predictions for $F_L$ 
with the data summarized in \cite{Taylor:2024pcf,Boroun:2025gig}, which were extracted from 
the HERA measurements of DIS cross section \cite{HERA}. The data for the proton structure 
function $F_L(x_B=Q^2/s,Q^2)$ were plotted as a function of $Q^2$ with $\sqrt{s}=300$ and
318 GeV in \cite{Taylor:2024pcf,Boroun:2025gig}. For $x_B=0.18$-0.65
considered in this work, the photon virtuality covers the high
$Q^2\sim 1.6\times 10^4$-$5.9\times 10^4$ GeV$^2$ for $\sqrt{s}=300$ GeV. Below we 
focus on the two typical points $x_B=0.25$ and $0.40$.

The perturbative input for the structure function $F_1$ up to NLO shares the similar 
expression to that for $F_2$ in Sec.~II, with the functions $S_1$, $V_1$, $N_1$ and 
$f_1^Q$, which are also referred to the Appendices of \cite{Kretzer:1998ju}, being 
substituted for $S_2$, $V_2$, $N_2$ and $f_2^Q$. We then combine the inputs for $F_2$ 
and $F_1$ into that for $F_L$. Iterating the same steps, and searching for the minimal 
difference $d$ by scanning over $\Lambda$, we identify the physical solution 
which describes $F_L$ from low to high baryon masses. Next we read off the proton
structure function $F_L(x_B,Q^2$) at the proton mass $m_p$. For $\sqrt{s}=300$ GeV, we have
\begin{eqnarray}
F_L(x_B=0.25,Q^2=0.25s)=0.15^{+0.01}_{-0.04},
\end{eqnarray}
where the central value (upper bound, lower bound) corresponds to $\Lambda=41000$ GeV
($\Lambda=32000$ GeV, $\Lambda=70000$ GeV) and $N=18$, and
\begin{eqnarray}
F_L(x_B=0.40,Q^2=0.40s)=0.13\pm 0.01,
\end{eqnarray}
where the central value (upper bound, lower bound) corresponds to $\Lambda=40000$ GeV
($\Lambda=35000$ GeV, $\Lambda=51000$ GeV) and $N=18$. 
We claim the proton structure function $F_L\approx 0.1$ for finite $x_B$, which
descends with $Q^2$, in consistency with the Callan-Gross relation. The data for $F_L$, 
located between 0 to 0.2 for $Q^>10^4$ GeV$^2$ with large uncertainties 
\cite{Taylor:2024pcf,Boroun:2025gig}, match our predictions.

\subsection{Pion Structure Function $F_2^\pi$}

The dispersive formalism can be applied to the study of the pion structure functions
straightforwardly as stated in the Introduction. Following the quark counting rules 
\cite{Brodsky:1973kr,Ball:2016spl}, we ought to take the number 
of spectator partons $n_s=1$ for a pion, which determines the threshold behavior of the 
structure function $F_2^\pi\sim 1-x_\pi$ as the Bjorken variable $x_\pi\to 1$, and the 
index $\alpha=1/2$ for the generalized Laguerre polynomials. Nevertheless, this setting 
yields negative $F_2^\pi$ at higher $Q^2\sim 10^3$ GeV$^2$, which violates positivity.
Therefore, we choose a bit larger $\alpha=1$ to cure this problem in view that the quark 
counting rules are not exact. The modifications for the pion case compared to the proton 
DIS in the previous subsections then include (1) the parameter $\gamma=2$ in the ultraviolet 
subtraction term accordingly in order not to impact the threshold 
behavior $F_2^\pi\sim (1-x_\pi)^2$; (2) the multiplication of the solution to the 
dispersion relation by the corresponding charge factor $(2/3)^2+(1/3)^2=5/9$ for a 
charged pion (the charged factor for a neutral pion is the same); (3) the replacement 
of the proton mass $m_p$ by the pion mass $m_\pi=0.14$ GeV \cite{PDG}, at which $F_2^\pi$ 
is read off. The perturbative input from the heavy quark regime is identical.
On the experimental side, the pion structure function $F_2^\pi(x_\pi,Q^2)$ has been 
deduced from the measurements of the neutron-tagged DIS cross section 
$\ell p\to  \ell' Xn$ up to the Bjorken variable $x_\pi\sim 0.12$ and the photon virtuality 
$Q^2\approx 10^3$ GeV$^2$ \cite{ZEUS:2002gig}.

To compare our predictions with the data, we work out 
\begin{eqnarray}
& &F_2^\pi(x_\pi=0.12,Q^2=120\;{\rm GeV}^2)\approx 0.12,\nonumber\\
& &F_2^\pi(x_\pi=0.12,Q^2=10^3\;{\rm GeV}^2)\approx 0.24,\label{p2}
\end{eqnarray}
which correspond to $\Lambda=250$ GeV, $N=14$ and $\Lambda=2100$ GeV, $N=15$,
respectively. The errors within $\pm 0.01$ are not given explicitly. 
One should bear in mind that gluonic parton contributions may not be negligible at the
Bjorken variable $x_\pi=0.12$, so substantial theoretical errors for Eq.~(\ref{p2})
are expected. We present more predictions for the confrontation with future data,
\begin{eqnarray}
& &F_2^\pi(x_\pi=0.25,Q^2=10^2\;{\rm GeV}^2)\approx 0.09,
\nonumber\\
%& &F_2^\pi(x_\pi=0.25,Q^2=10^3\;{\rm GeV}^2)\approx 0.21,
& &F_2^\pi(x_\pi=0.40,Q^2=10^3\;{\rm GeV}^2)\approx 0.17,
\label{p3}
\end{eqnarray}
which correspond to $\Lambda=440$ GeV, $N=15$ and $\Lambda=7100$ GeV, $N=16$, respectively.

The measurement of the pion structure function also suffers huge uncertainties originating 
from the normalization of the pion flux; the outcomes from the additive quark model could 
be greater than those from the effective one-pion-exchange flux (EF) model by a factor of 2. 
The predictions in Eq.~(\ref{p2}) are closer to the data from the latter 
(see Fig.~18 in \cite{ZEUS:2002gig}),
\begin{eqnarray}
& &F_2^{\pi,{\rm EF}}(x_\pi=0.12,Q^2=120\;{\rm GeV}^2)\approx 0.15,\nonumber\\
& &F_2^{\pi,{\rm EF}}(x_\pi=0.12,Q^2=10^3\;{\rm GeV}^2)\approx 0.18\pm 0.02.
\end{eqnarray}
As remarked in \cite{ZEUS:2002gig}, $F_2^\pi$ has the similar dependencies on the Bjorken 
variable and the photon virtuality as those of the proton structure function $F_2$. However,
our calculation of $F_2^\pi$ for higher $x_\pi$ seems not to endorse this observation.
The ratio $F_2^\pi/F_2$ takes the values
\begin{eqnarray}
& &\frac{F_2^\pi(x_\pi=0.25,Q^2=10^2\;{\rm GeV}^2)}
{F_2(x_B=0.25,Q^2=10^2\;{\rm GeV}^2)}=0.33,\nonumber\\
& &\frac{F_2^\pi(x_\pi=0.40,Q^2=10^3\;{\rm GeV}^2)}
{F_2(x_B=0.40,Q^2=10^3\;{\rm GeV}^2)}=1.7.\label{p4}
\end{eqnarray}
The former is indeed similar to the scaling factor 0.361 found in \cite{ZEUS:2002gig}, but
the latter from the kinematic region experiments have not yet probed deviates from it.
We point out that the latter agrees with the one inferred from the prediction 
for $F_2^\pi$ based on the factorization theorem \cite{P:2026crg} (see their Fig.~6)
and the data for the proton $F_2$ \cite{H1:2003xoe}.

\section{CONCLUSION}

We have proposed a novel approach to high-energy inclusive QCD processes, which differs
dramatically from the conventional factorization theorem in concept, and demonstrated its
success by taking the proton DIS structure functions as an example. It relies only on 
analyticity of physical observables, takes as inputs standard OPE under the short-distance 
factorization, and solves associated dispersion relations directly as an inverse problem. 
That is, we derive nonperturbative observables at low hadron masses from perturbative 
inputs in the heavy quark regime through their connection governed by analyticity. The 
idea is inspired by QCD sum rules, where dispersion relations are established for 
correlation functions formed by QCD vacua. Here we work on scattering amplitudes with 
specific external particles. Excited-state contributions, which are difficult to remove 
in sum rules, can be avoided trivially in our formulation. The  
integration along branching cuts on a complex mass plane involves nonperturbative dynamics 
in the low-mass region in a similar way. The framework is not constructed on the light 
cone, so it does not demand the introduction of PDFs. The necessary nonperturbative inputs 
can arise, instead, from the hadronic matrix elements of higher-dimensional local operators 
in power corrections to OPE and from hadron masses, at which observables 
are retrieved. 

We stress that the OPE for the heavy baryon DIS is reliable, and does not break
down even in the limit $x_B\to 1$; the effective Bjorken variable for the heavy baryon
DIS is always lower than $x_B$ for the proton DIS and away from the threshold. Hence, 
there is sufficient phase space for real radiative corrections to realize the infrared 
cancellation with virtual radiative corrections. The collinear divergences do not exist, 
since both initial and final quarks are massive. The reliable OPE inputs then warrant the 
sensible determination of physical quantities from dispersive analyses. Although PDFs 
are not requested, the predictions need prior information, which forces the threshold 
behavior of the structure functions alleged by the quark counting rules, and the scaling 
of the structure functions in the photon virtuality $Q^2$. The priors help pin down the
structure functions around the proton mass and at high masses up to order of the photon 
energy $q^0$. It was shown that convergent solutions could be obtained in the inverse 
matrix method before invoking ill-posedness of an inverse problem. The physical 
solutions are stable against the variation of the arbitrary scale $\Lambda$ coming along 
with the ultraviolet regularization of the dispersion relations.

We read the proton structure functions from the solutions by assigning the baryon mass 
to the proton mass, which reproduce the data in the considered $x_B$-$Q^2$ domain. We 
concentrated on the quark contributions, so the investigation is 
restricted to finite Bjorken variables. It has been corroborated that our results for the 
proton structure functions $F_2$ and $F_L$ agree with the data in the range $x_B=0.18$-0.65 
and $Q^2=10^2$-$10^4$ GeV$^2$ satisfactorily. The extension to the pion 
structure function also accommodates the data for $x_\pi=0.12$ and 
$Q^2\approx 10^2$-$10^3$ GeV$^2$. This work offers not only a completely distinct 
viewpoint but a deeper insight into inclusive QCD processes. Phenomenologically, it allows 
a theoretical access to the large $x_B$ region which is hard to explore experimentally. 
Then PDFs at large momentum fractions can be constrained from the derived structure 
functions by handling the relevant factorization formulas as another inverse problem. The 
solutions apply to DIS of heavier baryons with the same quantum number in 
principle; the corresponding predictions can be read off by simply tuning the baryon mass 
to those of heavier states. The data for the $\Lambda$ baryon DIS extracted from 
measurements of Sullivan processes in the future will test our proposal.

The current leading-twist setup ought to be refined and further developed certainly. 
Numerous interesting subjects, such as higher-order and high-powers effects in OPE, 
can be inspected, which may improve the consistency between the theory and the data.  
We will extend the analysis of the structure functions to lower $Q^2$ by including 
cat's ears diagrams, and compare predictions with the available data \cite{HERMES:2011yno}. 
Besides, how to evaluate gluonic parton and multi-parton contributions deserve attention. 
It is also important to introduce isospin breaking effects, which differentiate observables 
associated with a proton and a neutron, so that the ambiguous ratio $F_2^n/F_2$ can be 
clarified. Another crucial direction concerns the application to Drell-Yan processes, 
which will strengthen the potential of our formalism for analyzing various 
inclusive QCD processes.

\vskip 1.0cm
{\bf Acknowledgement}

This work was supported in part by the National Science and Technology Council of Republic 
of China under Grant No. NSTC-113-2112-M-001-024-MY3.

\end{document}